\documentclass[a4paper,twocolumn,english,aps,prl,superscriptaddress]{revtex4}
\newcommand{\Tr}{\mbox{Tr}}
\usepackage[T1]{fontenc}
\usepackage[latin1]{inputenc}
\usepackage{float}
\usepackage{graphicx,color}
\usepackage{bm}
\usepackage{bbm}
\usepackage{amsmath}
\usepackage{color}
\usepackage{amssymb}
\makeatletter

\usepackage{babel}
\makeatother
\begin{document}

\title{Rabi oscillations of two-photon states in nonlinear optical resonators }

\author{Y. Sherkunov}
\address{National Graphene Institute, University of Manchester, Manchester, M13 9PL, UK}
\author{David M. Whittaker}
\address{Department of Physics and Astronomy, University of Sheffield, Sheffield S3 7RH, UK}
\author{Vladimir Fal'ko}
\address{National Graphene Institute, University of Manchester, Manchester, M13 9PL, UK}

\begin{abstract}
We demonstrate that four-wave mixing processes in high-quality non-linear resonators can lead to Rabi-like oscillations in photon occupation numbers and second-order correlation functions, being   a characteristic feature of the presence of entangled photon pairs in the optical signal. In the case of a system driven by a continuous coherent pump, the oscillations occur in the transient regime. We show that driving the system with pulsed coherent pumping would  generate strongly anti-bunched photon states.  
\end{abstract}

\maketitle

\section{Introduction and model}
Four-wave mixing (FWM) is a reversible process, in which a pair of photons, \textit{e.g.}, with a frequency $\omega_0$ is converted into two photons with frequencies $\omega_1$ and $\omega_2$, and  $\omega_1+\omega_2=2\omega_0$. As a result, FWM  could be used (instead of conventional down-conversion processes \cite{Kwiat95,Carusotto99}) to create entangled photon-pairs \cite{Wang01,Li05,Fulconis07,Cohen09,Kumar14}, suitable  for practical implementation in quantum cryptography \cite{Gisin07}.

Here, we study FWM  in high-quality resonators with discrete spectra \cite{footnoteA} and strong Kerr non-linearity: "zero-dimensional" exciton-polariton microcavities   \cite{Daif06, Ferrier11, Azzini11, Dufferwiel14, Besga15} and microwave resonators coupled to  superconducting qubits \cite{Wallraff04,Bishop09, Devoret13, Barends14}.   In these systems the FWM between photon pairs is reversible, so that quantum oscillations can occur between the initial and final states.  We show that for a photon pair in a resonator, its states oscillate between two  entangled  two-photon configurations with the frequency determined by the strength of Kerr non-linearity. For a pumped system, such Rabi-type oscillations would occur in transient regime after the pump is switched on, until losses damp the oscillations establishing a steady state. We suggest that  such oscillations would be a characteristic feature of the presence of entangled photon pairs in the optical signal. We also demonstrate that applying pulsed driving to the system would generate photon states in the anti-bunching regime.

\begin{figure}[h]
\includegraphics[width=0.45\textwidth]{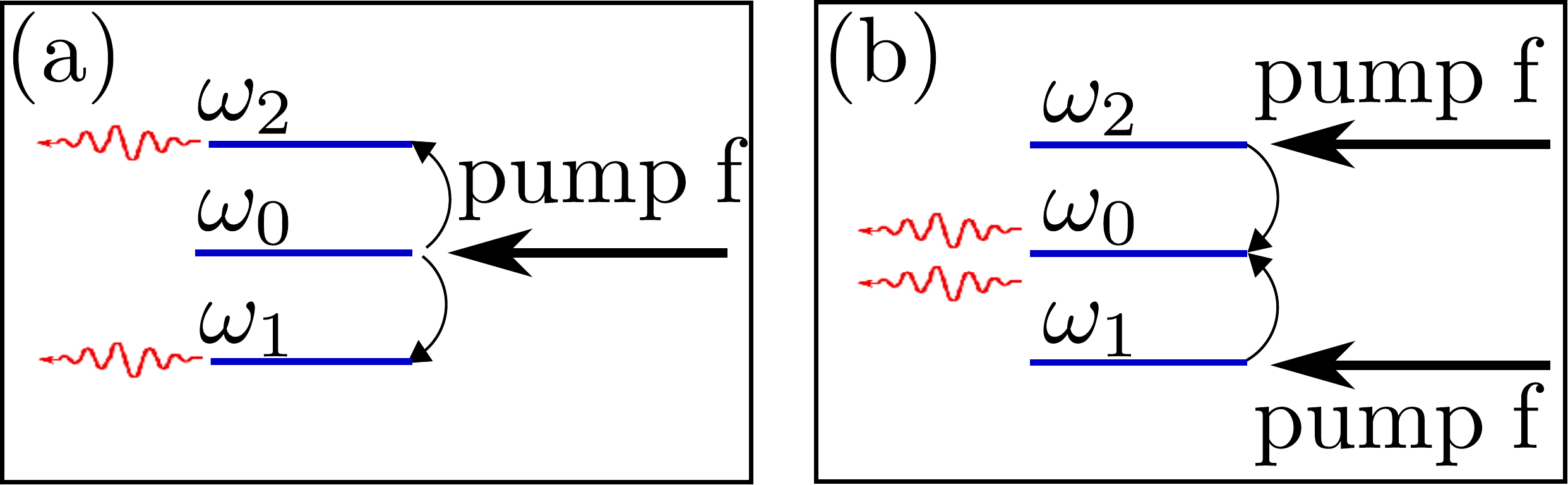}
\caption{Schematic diagram for a FWM process in a resonant microcavity with discrete spectrum. (a) Coherent  pump of $\omega_0$ mode with subsequent  scattering to the modes $\omega_1$ and $\omega_2$. (b) Coherent pump of two modes $\omega_1$ and $\omega_2$, generating a monochromatic pair in $\omega_0$ mode.     }  
\label{Fig1}
\end{figure}

To model  two-photon correlations in a non-linear microcavity, we employ the master equation approach to the density matrix of the photon system, $\rho$, \cite{Sherkunov14},
\begin{subequations}
\begin{eqnarray}
\frac{\partial\rho}{\partial t}&=&-i[H+H_{pi},\rho]\nonumber\\
&+&\gamma \sum_{i=0,1,2}(2 a_i\rho  a_i^{\dagger}-a_i^{\dagger}  a_i\rho -\rho  a_i^{\dagger} a_i)\label{mastereq},\\
H&=&\omega_1 a_1^{\dagger}a_1+\omega_2 a_2^{\dagger} a_2+\omega_0 a_0^{\dagger}a_0+H_{nl},\nonumber\\
H_{nl}&=&ua_1^{\dagger}a_2^{\dagger}a_0^2+h.c.. \label{H}
\end{eqnarray} \label{Hh}
\end{subequations} 
In (\ref{Hh}), the Hamiltonian $H$ ($\hbar=c=1$)  describes the conservative part of the photon system \cite{footnoteB},  $a_i$ and $a^{\dagger}_i$ are annihilation and creation operators of the photons in the mode $\omega_i$,  $u$ is a photon-photon coupling constant due to Kerr non-linearity, $\gamma$ is the photon decay rate, which is taken the same for all photon modes, and $H_{pi}$  describes the photon pump.  Below, we consider  microcavities excited using one of the two possible pumps sketched in Fig.\ref{Fig1}, 
\begin{subequations}
\begin{eqnarray}
H_{p1}&=&f(t)\left(e^{i\omega_{p0}t}  a_0 + e^{-i\omega_{p0}t} a_0^{\dagger}\right)\label{Hp1},\\
H_{p2}&=&f(t) \sum_{i=1,2}\left(e^{i\omega_{pi}t} a_i +e^{-i\omega_{pi}t} a_i^{\dagger}\right),\label{Hp2}
\end{eqnarray}\label{Hp}
\end{subequations}
where $f(t)$ and $\omega_{pi}$ are the amplitude and frequency of the pumping field. The pump (\ref{Hp1}) excites photons in the mode $\omega_0$, while pump (\ref{Hp2}) corresponds to coherent excitation of both  $\omega_1$ and $\omega_2$ modes. By solving  Eq (\ref{mastereq}) for the density matrix determined in the Fock space of  states with different photon numbers, $m_i$ and $n_i$, in the three modes,
\begin{equation}
 \rho=\sum \rho(m_1,m_2,m_0,n_1,n_2,n_0)|m_1,m_2,m_0\rangle \langle n_1,n_2,n_0|\label{rhoP},
\end{equation} 
we evaluate the occupation  numbers of photons and  zero time-delay pair correlation functions for each mode $\omega_i$,
\begin{eqnarray}
N_i&=&\Tr a_i^{\dagger}a_i\rho\label{Ni},\\
g^{(2)}_i&=&\Tr [(a_i^{\dagger})^2a_i^2]/N_i^2 \label{g2i}.
\end{eqnarray}
  
\begin{figure*}[ht]
\includegraphics[width=0.95\textwidth]{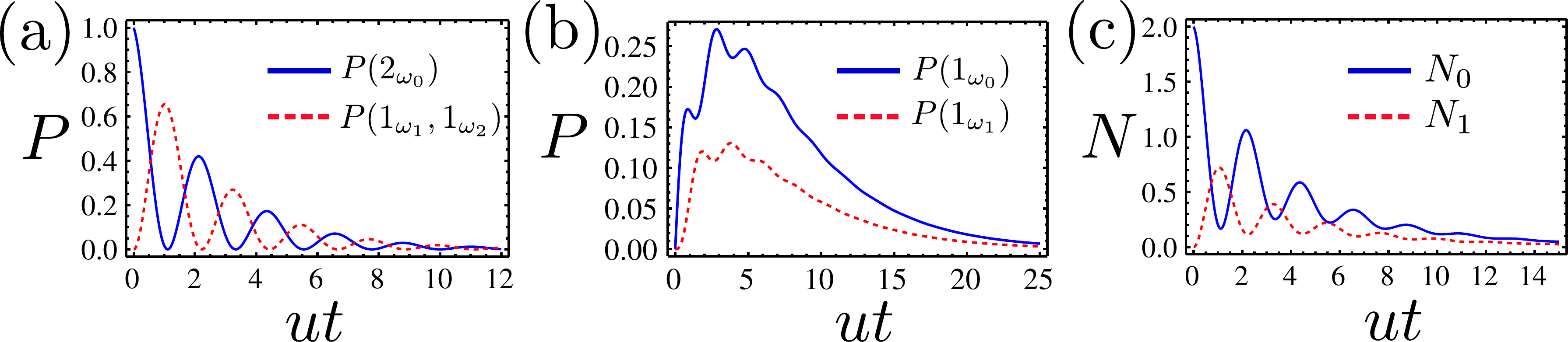}
\caption{Evolution of a two-photon state prepared in the mode $\omega_0$ in the resonator with damping $\gamma=0.1u$, $\delta=0$, and $f=0$. (a) Probability distribution of two-photon states. (b) Probability distribution of  single-photon states. (c) Average number of photons in the modes $\omega_0$ and $\omega_1$, or $\omega_2$. }
\label{Fig2}
\end{figure*} 

\section{Rabi oscillations of two-photon states in closed cavities}
To introduce the idea of Rabi oscillations of two-photon states, let us consider an initial state of the system with two $\omega_0$ photons in an isolated ($f(t)=0$) three-mode ($\omega_0$, $\omega_1$, $\omega_2$)  non-dissipative ($\gamma=0$)  cavity, with $\delta=\omega_1+\omega_2-2\omega_0\ll\omega_i$. The Hamiltonian  of this system (\ref{H}) can be projected onto the reduced Fock space of the two-photon states, $\{|2_{\omega_0}\rangle,|1_{\omega_1},1_{\omega_2}\rangle\}$, \[H= \left( \begin{array}{cc}
2\omega_0 & \sqrt 2 u  \\
\sqrt 2 u & \omega_1+\omega_2 \end{array} \right),\] with the eigenvalues and eigenstates,
\begin{subequations}
\begin{eqnarray}
E_{\pm}=\omega_0+(\omega_1+\omega_2)/2\pm \Omega/2,\;\;\Omega=\sqrt{\delta^2+8u^2},\label{Enondiss}\\
\psi_{\pm}=\frac{\pm 1}{2\sqrt{\Omega}}\left(\sqrt{\Omega\mp\delta}( a_0^{\dagger})^2\pm\frac{4ua_1^{\dagger} a_2^{\dagger}}{\sqrt{\Omega\mp\delta}} \right)|0\rangle.\label{psinondiss}
\end{eqnarray}
\end{subequations}
Here, $\Omega(\delta\rightarrow 0)\rightarrow 2\sqrt 2 u$ is the  two-photon Rabi frequency, which characterises time-dependent oscillations of the probabilities $P(2_{\omega_0})$ and $P(1_{\omega_1},1_{\omega_2})$ to find these two photons, $$P(2_{\omega_0})=\cos^2\left(\frac{\Omega t}{2}\right)+\frac{\delta^2}{\Omega^2}\sin^2\left(\frac{\Omega t}{2}\right),$$ $$P(1_{\omega_1},1_{\omega_2})=\frac{8u^2}{\Omega^2}\sin^2\left(\frac{\Omega t}{2}\right).$$

\section{Rabi oscillations in open resonators}
In the case of a lossy system, with a finite decay rate $\gamma$, this result can be generalised into the formulae for probabilities to find two or one photon in the cavity, illustrated in Figs. \ref{Fig2}(a) and (b),   
\begin{widetext}
\begin{eqnarray}
& & P(2_{\omega_0})=e^{-4\gamma t}\left[\cos^2\left(\frac{\Omega t}{2}\right)+\frac{\delta^2}{\Omega^2}\sin^2\left(\frac{\Omega t}{2}\right)\right],\;\; P(1_{\omega_1},1_{\omega_2})=e^{-4\gamma t}\frac{8u^2}{\Omega^2}\sin^2\left(\frac{\Omega t}{2}\right),\nonumber\\
& & P(1_{\omega_0})=\frac{2e^{-4\gamma t}}{\Omega^2(\Omega^2+4\gamma^2)}\left[\left(e^{2\gamma t}-1\right)(\Omega^2-4u^2+4\gamma^2)\Omega^2+16\gamma u^2\sin\left(\frac{\Omega t}{2}\right)\left(\Omega\cos\left(\frac{\Omega t}{2}\right)+2\gamma\sin\left(\frac{\Omega t}{2}\right)\right)\right],\nonumber\\
& & P(1_{\omega_1})=P(1_{\omega_2})=\frac{4e^{-4\gamma t}u^2}{\Omega^2(\Omega^2+4\gamma^2)}\left[\Omega^2\left(e^{2\gamma t}-1\right)-4\gamma\sin\left(\frac{\Omega t}{2}\right)\left(\Omega\cos\left(\frac{\Omega t}{2}\right)+2\gamma \sin\left(\frac{\Omega t}{2}\right)\right)\right].\label{p1w1}
\end{eqnarray}

These oscillations can manifest themselves in the temporal evolution of the occupation numbers, $N_i$, illustrated in Fig. \ref{Fig2}(c),

\begin{eqnarray}
N_0&=&\frac{2e^{-4\gamma t}}{(4\gamma^2+\Omega^2)\Omega}\left[e^{2\gamma t}\Omega(4\gamma^2+\Omega^2-4u^2)+4u^2(\Omega\cos(\Omega t)+2\gamma \sin(\Omega t))\right],\nonumber\\
N_1&=&N_2=\frac{4u^2e^{-4\gamma t}}{(4\gamma^2+\Omega^2)\Omega}\left[e^{2\gamma t}\Omega-(\Omega\cos(\Omega t)+2\gamma \sin(\Omega t))\right].\label{Rabi}
\end{eqnarray}
\end{widetext}

In principle, the quantum-optical Rabi-type oscillations can occur in systems driven by coherent pumping. We demonstrate this  for the  case of resonant pumping, $\omega_{pi}=\omega_i$, and  $\delta=0$. In the case of lossless system  ($\gamma=0$) and weak continuous pumping switched on at $t=0$, $f(t)=(const\ll u)\times \theta(t)$, where $\theta(t)$ is the Heaviside function, Eq. (\ref{mastereq}) can be solved analytically by diagonalizing the Hamiltonian (\ref{H}) and (\ref{Hp}), which in the interaction representation, can be written in the time-independent basis. This procedure leads to the following eigenvalues, $\{\pm f,\pm \sqrt 2 u\}$ for pump (\ref{Hp1}), and $\{0,\pm\sqrt{2}f,\pm\sqrt 2u\}$  for pump (\ref{Hp2}), and for $ft\ll 1$ results in the oscillations in occupation numbers ($\delta =0$),
\begin{eqnarray}
N_1(t)=N_2(t)=\frac{f^4 t^2}{u^2}\left(1-\frac{\sin\left(\Omega t/2\right)}{\sqrt 2ut}\right)^2\label{N1}
\end{eqnarray}
for pump (\ref{Hp1}), and
 \begin{eqnarray}
N_0(t)=\frac{4f^4 t^2}{u^2}\left(1-\frac{\sin\left(\Omega t/2\right)}{\sqrt 2ut}\right)^2\label{N0}
\end{eqnarray} 
for pump (\ref{Hp2}). Here, the sizeable part of the oscillations has frequency $\Omega/2$.

\begin{figure}[ht]
\includegraphics[width=0.48\textwidth]{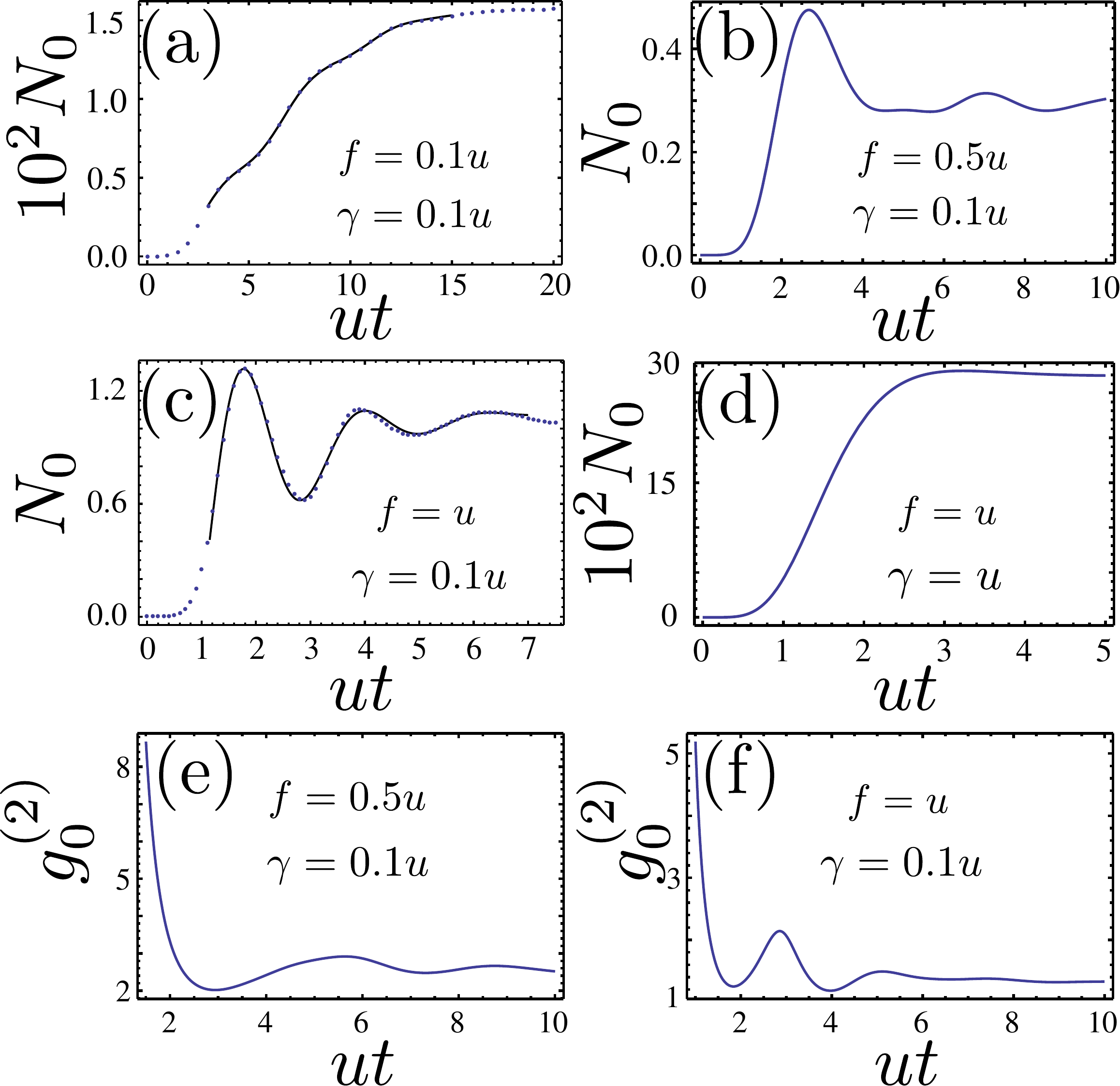}
\caption{Numerical results for occupation numbers in the mode $\omega_0$, $N_0(t)$, and two-photon correlation function for the mode $\omega_0$, $g^{(2)}_0$,  in the continuously driven system sketched in Fig \ref{Fig1} (b). In (a) and (c),  dots represent numerical results fitted by Eq. (\ref{fit}) (solid line). The fitting parameters are: (a) $b_1=0.383$, $\alpha_1=3\gamma$, $\phi_1=1.361$, $b_2=0$ ; (c) $b_1=0$, $b_2=3.023$, $\alpha_2=8\gamma$, $\phi_2=1.137$.}
\label{Fig3}
\end{figure}
\begin{figure}
\includegraphics[width=0.48\textwidth]{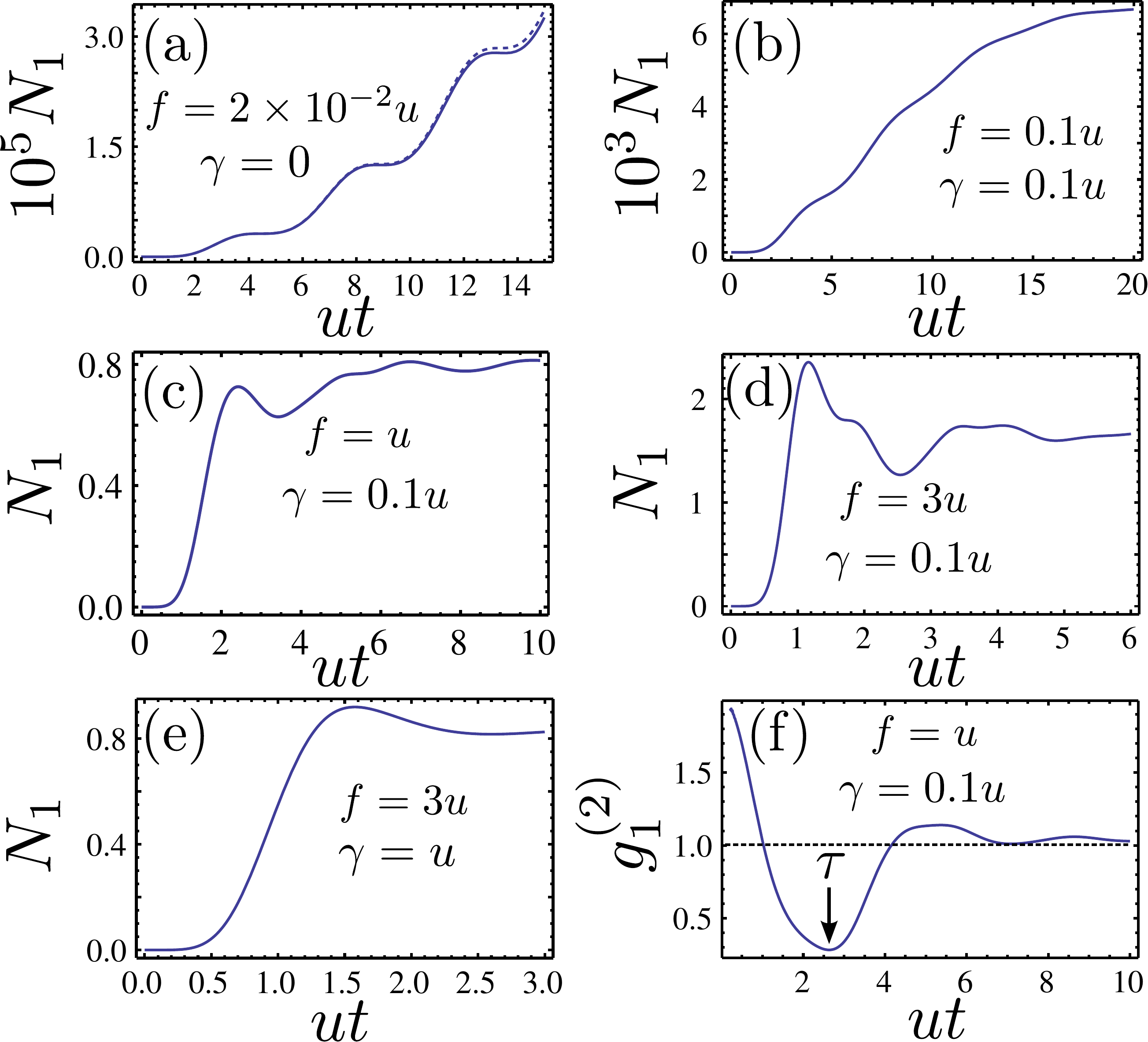}
\caption{Numerically evaluated occupation number of photons in the mode $\omega_1$, $N_1(t)$, and  two-photon correlation function in the mode $\omega_1$, $g^{(2)}_1$, for the continuously driven system shown in Fig. \ref{Fig1} (a). In (f), the arrow indicates the first minimum of $g^{(2)}_1$ at $\tau\approx 2.6 /u$.  In (a), we compare the results of numerical simulations (solid line) and Eq.(\ref{N1}) (dashed line). }
\label{Fig4}
\end{figure} 

For a larger pumping amplitude, $f=(const\sim u)\times\theta (t)$, and for finite losses $\gamma\neq 0$, we solve Eq (\ref{mastereq}) numerically.  We truncate the Fock space  photon numbers,  $m_i$, $n_i\leq N_{\max}$ and check that, for $f\sim u$,  $N_{max}=10$ is enough for the convergence (upon the increase of $N_{max}$) of the numerical results. We find that  photon pairs manifest themselves as Rabi-like oscillations between the two states $|2_{\omega_0}\rangle$ and $|1_{\omega_1},1_{\omega_2}\rangle$ as we show in Fig.\ref{Fig3} for the pumping scheme of Fig. \ref{Fig1}(b).  Interestingly, the oscillations of the occupation numbers can be well described by the two harmonics with frequencies close to $\Omega$ and $\Omega/2$, as in the case discussed above, and can be fitted by the interpolation formula (Fig.\ref{Fig3}(a), (c)), 

\begin{eqnarray}
N_0 &=& F(t)\left [1+b_1e^{-\alpha_1 t}\cos\left(\frac{\Omega t}{2}+ \phi_1\right)\right.\nonumber\\
&+&\left.b_2e^{-\alpha_2 t}\cos\left(\Omega t+\phi_2\right)\right],\label{fit}
\end{eqnarray} 
where $F(t)$ is a monotonous (non-oscillating) function. For weak pumping, $f\ll u$, the amplitude of the $\Omega/2$-harmonic, $b_1$, is much larger than the amplitude of the $\Omega$-harmonic, $b_2$, which is noticeable  in  Figs. \ref{Fig3}(a).   As the pumping amplitude $f$ increases, $b_2$ goes up and exceeds $b_1$ at $f\sim u$. Thus, the oscillations at $f\geq u$ have frequency   $\Omega$ (Fig. \ref{Fig3} (b)-(c)).  The oscillations with the same frequencies can also be observed in the two-photon  correlation function  $g^{(2)}_0$, Figs. \ref{Fig3}(e)-(f).

\begin{figure}
\includegraphics[width=0.48\textwidth]{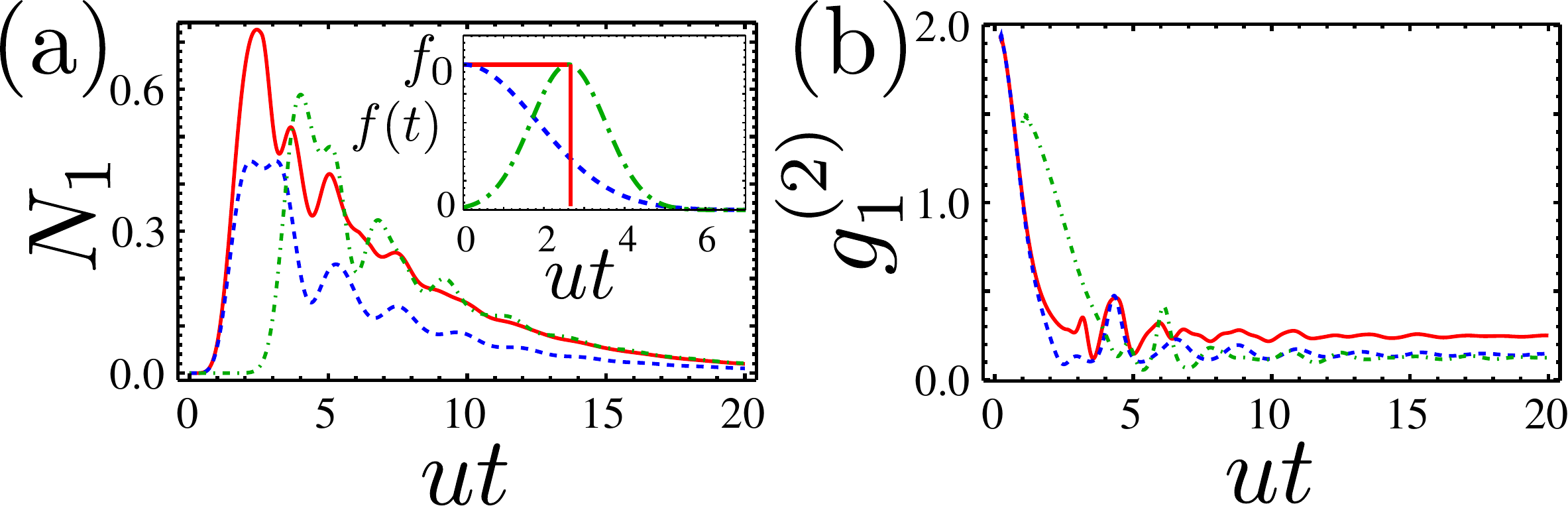}
\caption{Numerically evaluated  $N_1(t)$ and $g^{(2)}_1$ for the system of  Fig. \ref{Fig1} (a) with $\gamma=0.1u$ driven by  the pulses (see inset): $f(t)=f_0\theta(t)\theta(\tau-t)$ (red solid line); $f(t)=f_0\theta(t) \exp(-t^2/\tau^2)$ (dashed blue line); $f(t)=f_0\exp[-4(t-\tau)^2/\tau^2]$ (green dot-dashed line) with amplitude $f_0=u$ and  $\tau=2.6/u$, corresponding to the minimum of $g_1^{(2)}$ of a continuously pumped system (Fig.\ref{Fig4}(f)). }
\label{Fig5}
\end{figure}

We find a similar behaviour for  the pumping scheme illustrated in Fig. \ref{Fig1}(a). In Figs. \ref{Fig4}(a)-(d), the  oscillation frequency changes from $\Omega/2$ for  $f=(const\ll u)\times \theta(t)$ to $\Omega$ for $f=(const\sim u)\times\theta (t)$. There are also oscillations in  the two-photon correlation function in the mode $\omega_1$,  $g^{(2)}_1$,  with the period matching the one of $N_1$. Note that, during its cycle, $g^{(2)}_1$ crosses unit value, indicating switching  from  bunching  to  anti-bunching. We note that anti-bunching can be promoted, if the system is driven by pulses with a pump of finite duration $\tau$ corresponding to the minimum of  $g^{(2)}_1$ in a continuously  pumped system. As we show in Fig.\ref{Fig5} for  the cavity driven by $\omega_0$-mode pulses, $f(t)=f_0\theta(\tau-t)\theta(t)$, $g^{(2)}_1$ has a maximum corresponding to the bunching regime at the time when $N_1$ is small, but then, the system switches into  anti-bunching regime. In principle, the form of the ringing tones in oscillations of $N_1$ and $g^{(2)}_1$ may depend on the extent and shape of the excitation pulse, however this dependence is weak, as we show in Fig.\ref{Fig5}, where we compare the systems excited by the pulses of various shapes.       
   
\section{Discussion}
In this paper, we have demonstrated that in non-linear resonators with discrete spectrum correlated photon pairs can manifest themselves as quantum Rabi-like oscillations with the period determined by the strength of non-linearity. In coherently driven systems, these oscillations would occur in the transient regime, $t<1/\gamma$, as shown in Figs. \ref{Fig3} and Fig. \ref{Fig4}, thus, they could be observed in a system with  sufficiently low damping, $\gamma\leq u$.   To mention, the proposed theory describes  systems with various frequency ranges. It can be applied to the visible-range polaritonic microcavities based on GaAs, where the non-linearity is due to exciton-exciton interaction in GaAs \cite{Timofeev14}. In these systems, the strength of non-linearity, $u \sim 10\mu eV$, is still too weak as compared to damping, $\gamma\sim 80\mu eV$  \cite{Dufferwiel14}. However, we suggest that use of  two-dimensional transition metal dichalcogenides \cite{Britnell13,Palummo15,Schwarz14} in microcavities, may improve the $u/\gamma$ ratio due to stronger exciton-exciton interaction.  Moreover, the proposed theory of two-photon oscillations is also applicable to microwave photons in superconducting resonators, where one can achieve quality factors in the range of $10^6 - 10^9$ \cite{Barends10, Geerlings12, Reagor13}, hence reach the regime $u\gg \gamma$.

\begin{figure}
\includegraphics[width=0.48\textwidth]{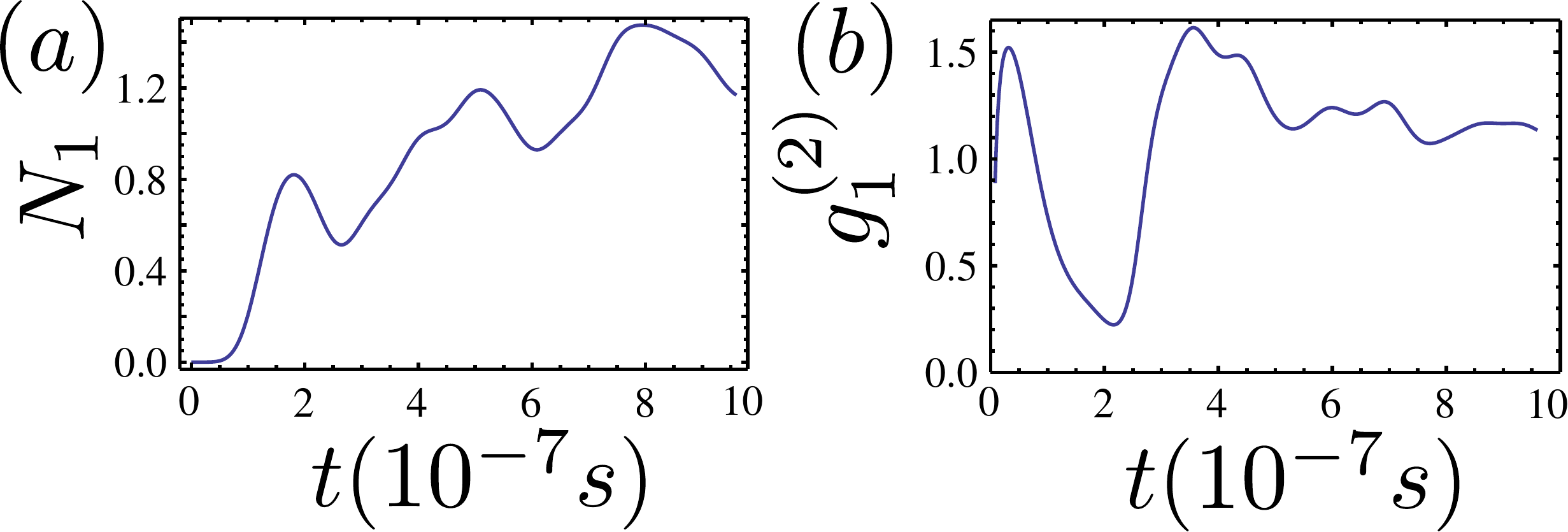}
\caption{Numerically evaluated (a) $N_1(t)$ and (b) $g^{(2)}_1$ for a high-quality toroidal microcavity coupled to atoms in an EIT regime driven by resonant coherent pumping of Fig. \ref{Fig1}(a) for an experimentally achievable set of parameters: $u=1.25\times 10^7s^{-1}$, $\gamma=2\times 10^5 s^{-1}$, and $f=1.25\times 10^7s^{-1}$ \cite{Armani03,Aoki06,Spillane05,Hartmann07}.}
\label{Fig6}
\end{figure}

Large $u/\gamma$ ratio can also be  achieved by resonantly coupling atoms with level structure allowing for electromagnetically induced transparency (EIT)  \cite{Imamo97, Hartmann07} and  microcavities with high quality factors, $Q$, such as toroidal ($Q>10^8$) \cite{Armani03} or microrod ($Q>10^9$)  \cite{Del13} resonators. In these systems, the strength of non-linearity can reach $u\sim 1.25 \times 10^7 s^{-1}$, while the losses can be as low as $\gamma\sim 2 \times 10^{5}s^{-1}$ \cite{Armani03,Aoki06,Spillane05,Hartmann07}, hence reaching the desirable regime $u\gg\gamma$ leading, as we show in Fig.\ref{Fig6}, to Rabi oscillations in photon occupation numbers and two-photon correlation function.

We have also shown, that application of pulsed pumping to the system could lead to the generation of squeezed states of strongly anti-bunched photons.

We thank D. Krizhanovskii, E. Cancellieri and M. Skolnick for useful discussions. This
work was supported by EPSRC Programme Grant
EP/J007544.

\bibliography{ybs.bib}{}

\begin{thebibliography}{35}
\bibitem[{foo({\natexlab{a}})}]{footnoteA}
\bibinfo{note}{By now, the studies of FWM processes have been focused on
  quasi-one-dimensional systems, such as optical fibres \cite{Vered15}.}
\bibitem[{foo({\natexlab{b}})}]{footnoteB}
\bibinfo{note}{In our analysis we neglect the effects of self- and cross phase
  modulation, which occur in a system with Kerr non-linearity
  \cite{MandelWolf}, although, we do not expect that these will affect the
  frequency of the oscillations but may enhance damping.}
\end{thebibliography}
\bibliographystyle{apsrev}
\end{document}